\documentclass[twocolumn,superscriptaddress,aps,prl]{revtex4-1}
\usepackage{graphicx,amsmath,natbib}

\begin{document}

\title{Terahertz nonlinear conduction and absorption saturation in silicon waveguides}

\author{Shanshan Li}
\email{lsscat@umd.edu}
\affiliation{Institute for Research in Electronics \& Applied Physics, University of Maryland, College Park, MD 20742, USA}
\author{Gagan Kumar}
\email{gk@iitg.ernet.in}
\affiliation{Department of Physics, Indian Institute of Technology Guwahati, Guwahati, 781039, India}
\author{Thomas E. Murphy}
\email{tem@umd.edu}
\affiliation{Institute for Research in Electronics \& Applied Physics, University of Maryland, College Park, MD 20742, USA}

\begin{abstract}
We employ a silicon dielectric waveguide to confine and concentrate terahertz pulses, and observe that the absorption saturates under strong terahertz fields.  By comparing the response between lightly-doped and intrinsic silicon waveguides, we confirm the role of hot carriers in this saturable absorption.  We introduce a nonlinear dynamical model of Drude conductivity that, when incorporated into a wave propagation equation, accurately reproduces the observations and elucidates the physical mechanisms underlying this nonlinear effect.  The results are numerically confirmed by Monte Carlo simulations of the Boltzmann transport equation, coupled with split-step nonlinear wave propagation.
\end{abstract}%

\maketitle

Among semiconductors, silicon is not only the most prevalent material in electronics, but it is also one of the most favorable dielectric materials for terahertz applications.  Intrinsic silicon is transparent at wavelengths longer than 1100 nm, and has exceptionally low loss in the far infrared\cite{grischkowsky1990,*dai2004,*yee2009}.  While the nonlinear properties and applications of silicon are well established in the near-infrared and mid-infrared regime\cite{rong2005,*foster2006,*lin2007,*zhang2007,*jalali2010,*kuyken2011}, there have been very few observations of nonlinear propagation in the terahertz regime.

The terahertz photon energy (4.1 meV at 1 THz) is too small to produce new carriers in silicon through a 1- or 2-photon absorption, and hence the linear and nonlinear properties are caused by acceleration or heating of the existing electron (or hole) population.  The traditional Drude model of conductivity that is commonly used to describe free carrier absorption and dispersion in silicon in the terahertz regime fails to explain nonlinear wave propagation effects.

In 2010, Hebling et al.\ and Kaur et al.\ independently observed THz field induced absorption bleaching in n-doped bulk silicon, using terahertz pump-probe measurements\cite{hebling2010} and z-scan measurements\cite{kaur2010}.  They suggested that the effect might be explained by scattering of electrons into a higher energy (L) valley within the conduction band.  Terahertz induced nonlinear effects have also been observed in a variety of other bulk semiconductors, including Ge\cite{hebling2010}, GaAs\cite{gaal2006,*hoffmann2009,*sharma2012,*turchinovich2012,*jeong2013}, GaP\cite{hoffmann2010} and InSb\cite{wen2008,*hoffmann2009b,*junginger2012}, and numerous hot carrier effects have been offered as explanations, including intervalley scattering, band nonparabolicity, and impact ionization.  In most cases, the observations were carried out using wafers or windows with optical thickness of only a few terahertz wavelengths.  In such thin samples, the cumulative nonlinearity is necessarily quite small, and it is difficult to separate propagation effects from interface effects such as small changes in reflectivity, or spatial effects such as self-focusing and diffraction.

To overcome these limitations, we couple picosecond terahertz pulses into a 2 cm long silicon dielectric ridge waveguide.  The waveguide greatly enhances the field concentration and nonlinear propagation length, thereby ensuring that the measured effect represents a true nonlinear wave interaction accumulated over hundreds of terahertz wavelengths, and also allows for interplay between the linear mode propagation and nonlinearity.  The waveguide configuration also eliminates spatial nonlinear effects like self-focusing, enabling unambiguous measurement of the temporal nonlinear behavior.  We observe over a two-fold increase in the power transmission ratio at high powers relative to low powers, depending on the carrier concentration, and we present a new physical and numerical model that explains the observed behavior.

\begin{figure}[b]
\centering
\includegraphics{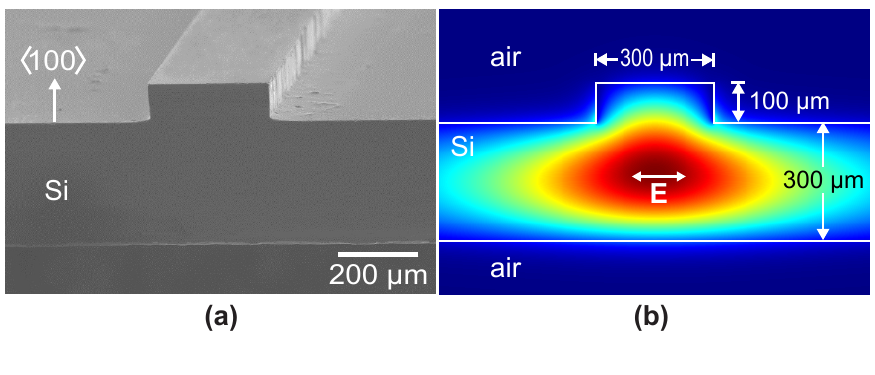}
\caption{(a) Cross-sectional micrograph of fabricated silicon ridge waveguide and (b) calculated TE eigenmode at 0.5 THz.}
\label{fig:1}
\end{figure}

The silicon ridge waveguides were fabricated from 400 $\mu$m thick, double-side polished (DSP), $\left<100\right>$ silicon wafers.  In order to better assess the role of carriers, we used two types of silicon:  lightly p-doped wafers with a nominal resistivity of of 150-350 $\Omega\cdot$cm and float-zone semi-insulating wafers with a resistivity of 10 k$\Omega\cdot$cm.  A 1 $\mu$m sacrificial layer of SiO$_2$ was deposited by CVD on the wafers, and patterned using contact photolithography and reactive-ion etching to produce a 300 $\mu$m wide oxide hard-mask for subsequent etching of the waveguides.  The waveguides were etched to a depth of 100 $\mu$m using pulsed deep reactive ion etching (Bosch process), after which the remaining photoresist and oxide hard mask were removed.  Fig.~\ref{fig:1}(a) shows a cross-sectional micrograph of the completed ridge waveguide, and Fig.~\ref{fig:1}(b) shows the corresponding fundamental TE eigenmode of the waveguide, calculated at 0.5 THz.  The transverse waveguide dimensions were chosen to ensure single-mode operation over the frequency range of interest.  The waveguides were cut to a length of 2 cm using a dicing saw.

\begin{figure}[tbp]
\centering
\includegraphics{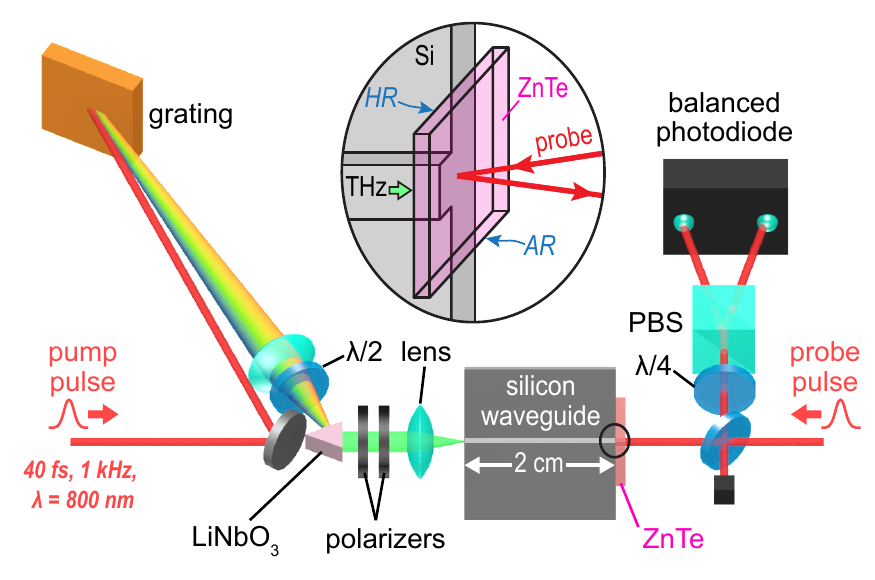}
\caption{Experimental setup used to measure the THz nonlinear transmission through the silicon waveguide.}
\label{fig:2}

\end{figure}

Fig.~\ref{fig:2} illustrates the experimental setup used to characterize the THz nonlinear response.  An amplified Ti-sapphire laser system produces 40 fs, 1 kHz repetition rate pulses at 800 nm center wavelength. The optical pulses are split (80:20) into pump and probe beams that are used for terahertz generation and detection, respectively.  The pump pulse impinges on a grating (2000 lines/mm), producing a $-1$ order diffracted beam that has a tilted pulse front\cite{hebling2002,*yeh2007,*hebling2008}.  The tilted pulse was de-magnified by a factor of 2$\times$ using a 60 mm focal length lens into a LiNbO$_3$ prism.  A $\lambda$/2 waveplate rotates the optical beam polarization from horizontal to vertical direction to align with the optical axis of the LiNbO$_3$.  The power of the THz output beam was adjusted using a pair of wire-grid polarizers, and focused using a polymethylpentene (TPX) lens onto the input waveguide facet.  The THz beam was linearly polarized in the $\left<011\right>$ crystallographic direction of the silicon waveguide.  Using the experimentally measured energy, pulse duration, and focused spot size of the terahertz beam, the peak electric field at the focus before inserting the waveguide was estimated to be 200 kV/cm\cite{yeh2007}.

The THz pulses impinging on and emerging from the waveguides were measured using both a pyroelectric detector and electrooptic sampling.  In the latter case, we used a 1 mm thick $\left<110\right>$ ZnTe crystal that was coated with an 800 nm dielectric mirror front face, and antireflection coating on the rear face, which allows the probe beam to be introduced in a reflection geometry\cite{seo2007,*chakkittakandy2008}, as shown in Fig.~\ref{fig:2}.  The ZnTe electrooptic crystal was placed in contact with the output facet of the waveguide, to allow for near-field optical sampling of the mode emerging from the waveguide.

To measure the nonlinear transmission through the waveguide, we used the Fourier transform to calculate the spectrum of the emerging waveform, and integrated the intensity spectrum to obtain a measure of the transmitted power.  For the range of powers considered, the nonlinearity of the electrooptic detection process was confirmed to be negligible in comparison to the absorption saturation in the silicon waveguide.

\begin{figure}[tbp]
\centering
\includegraphics{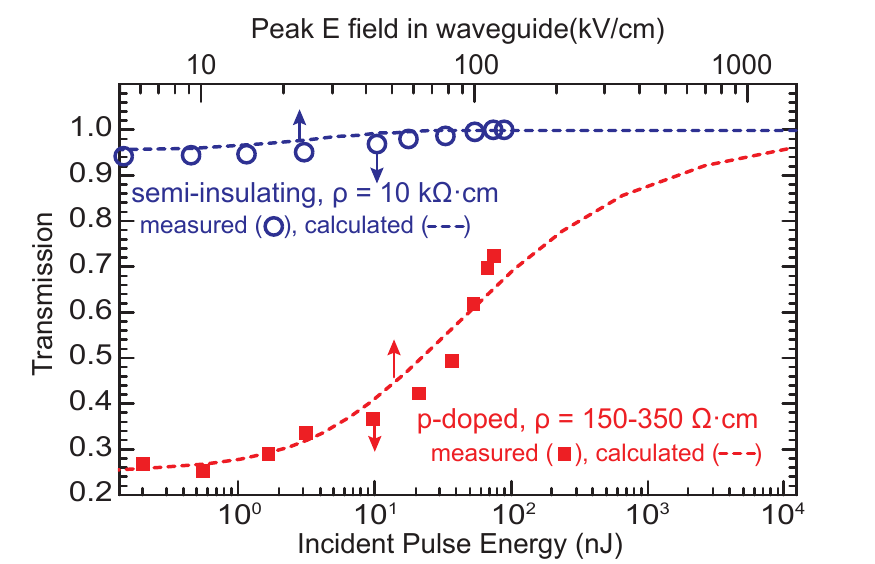}
\caption{Normalized power transmission for semi-insulating (circle) and doped (square) waveguides, and corresponding calculated (dashed lines) pulse energy transmission.}
\label{fig:3}
\end{figure}

Fig.~\ref{fig:3} shows the normalized transmission ratio as a function of the input pulse energy and peak field for the two waveguides considered here.  The semi-insulating silicon waveguide shows a small, but clearly measurable increase of 5\% in transmission as the pulse energy is increased from 0 to 75 nJ.  The p-type silicon waveguide, by contrast, shows a more than 2-fold increase in transmission at higher fluence.  The dashed lines plot the numerically calculated result (to be explained below), which shows that at sufficiently high pulse power, the power transmission ratio saturates at a level close to unity.  The fact that the saturable absorption is much stronger in doped silicon clearly demonstrates the role of free carriers in the nonlinear response.

A complete model of absorption in silicon waveguides must account for not only the field-dependent nonlinear carrier dynamics, but also the linear dispersion, which diminishes the peak field of the signal.  The terahertz nonlinear wave propagation can be described by a simplified one-dimensional wave equation,
\begin{equation}\label{eq:1}
  \left[\frac{\partial^2}{\partial z^2} - \frac{1}{c^2}\frac{\partial^2}{\partial t^2} \right]E = \mu_0 \left[\frac{\partial^2}{\partial t^2} P + \frac{\partial}{\partial t}J \right]\,,
\end{equation}
where $J$ is the current density (which is non-linearly related to $E$) and $P$ is the linear polarization of the material, which is linearly related to the electric field in the frequency domain by:
\begin{equation}\label{eq:2}
  \hat P(z,\omega)  = \epsilon_0 \left[n^2(\omega) - 1\right]\hat E(z,\omega)\,,
\end{equation}
where $n(\omega)$ is the effective the refractive index of the waveguide.

If the current is neglected, the forward traveling solution to \eqref{eq:1} in the frequency domain is
\begin{equation}\label{eq:3}
  \hat E(\Delta z,\omega) = \hat E(0,\omega) \exp \left[i\frac{\omega}{c}n(\omega)\Delta z\right]\,,
\end{equation}
where the refractive index $n(\omega)$ incorporates material and modal dispersion of the waveguide.

Conversely, if the dispersion is neglected but the current term is retained, then the wave equation can be written as:
\begin{equation}\label{eq:4}
  \frac{\partial^2E}{\partial z^2} - \frac{1}{\bar v^2}\frac{\partial^2E}{\partial t^2} = \mu_0\frac{\partial J}{\partial t}\,,
\end{equation}
where $\bar v \equiv c/n(\bar\omega)$ represents the average velocity of the terahertz pulse, evaluated at the center frequency of the spectrum.

Because the electric field travels in the $+z$ direction with an average velocity of $\bar v$, we assume that the resulting current density can be likewise cast as a function of a single argument, $J(t-z/\bar v)$, in which case \eqref{eq:4} can be integrated to find the field at the end of one step $\Delta z$:
\begin{multline}\label{eq:5}
  E(\Delta z,t) = E(0,t-\Delta z/\bar v) \\+\:\frac{\bar v^2\mu_0}{4}\int\limits_{t - 3\Delta z/\bar v}^{t - \Delta z/v}\left[J(t')-J(t - \Delta z/\bar v)\right]dt'
\end{multline}
The second term in \eqref{eq:5} represents a perturbation $\Delta E$ in the electric field caused by the current $J$.  The split-step numerical method replaces this accumulated nonlinearity by an equivalent lumped effect at $z=0$, which is found by advancing \eqref{eq:5} by the propagation time $\Delta z/\bar v$,
\begin{equation}\label{eq:6}
  \Delta E(t) = \frac{\bar v^2\mu_0}{4}\int\limits_{t - 2\Delta z/\bar v}^{t}\left[J(t')-J(t)\right]dt'
\end{equation}

The nonlinear wave propagation is numerically simulated by dividing the total propagation distance into steps of size $\Delta z$, computing the linear propagation for each increment in the Fourier domain using \eqref{eq:3}, and incorporating the nonlinearity as lumped in the time domain using \eqref{eq:6}.

The nonlinear relationship between the electric field $E(t)$ and current density $J(t)$ can be described using the balance equations obtained from the Boltzmann transport equations.  In the spatially homogeneous limit, the momentum balance equation is \cite{nougier1981}:
\begin{equation}\label{eq:7}
  \frac{dv}{dt} + \Gamma_m(\varepsilon) v = \frac{qE}{m^*}\,,
\end{equation}
where $v$ represents the carrier velocity, which is directly proportional to the current density through $J = Nqv$, and $\Gamma_m(\varepsilon)$ is the momentum relaxation rate, which we take to be a function of the energy, $\varepsilon$.

The energy balance equation is
\begin{equation}\label{eq:8}
  \frac{d\varepsilon}{dt} + \Gamma_\varepsilon \varepsilon = qEv\,,
\end{equation}
where $\varepsilon$ is the carrier energy relative to thermal equilibrium, and $\Gamma_\varepsilon$ is the energy relaxation rate.  The momentum and energy scattering rates are, in general, energy dependent, which couples these two equations.  We adopt the simple, and widely used model where the energy relaxation rate $\Gamma_\varepsilon$ is taken to be constant, while the momentum relaxation rate increases linearly with the carrier energy\cite{hansch1986}:
\begin{equation}\label{eq:9}
  \Gamma_m(\varepsilon) = \Gamma_{0} + \frac{\Gamma_\varepsilon \varepsilon}{m^*v_{\rm sat}^2}
\end{equation}
For sufficiently small carrier energy, the second term in \eqref{eq:9} may be neglected, in which case \eqref{eq:7} can be solved directly to give the familiar linear Drude relationship between $v$ and $E$, in the frequency domain,
\begin{equation}\label{eq:10}
  \hat v(\omega) = \frac{\mu}{1 -i\omega/\Gamma_0}\hat E(\omega)\,,
\end{equation}
where $\mu \equiv q/m^*\Gamma_0$ is the low-field mobility.  However, for sufficiently high fields, Eqs.~\eqref{eq:7}-\eqref{eq:9} predict well-known nonlinear transport phenomena including the saturation of carrier velocity at $v = v_{\rm sat}$ with increasing DC field strength.  The electron and hole saturation velocities in silicon are approximately $v_{\rm sat} \sim 10^7$ cm/s, and the corresponding critical electric field strength above which saturation effects become important is $E_{\rm cr} = v_{\rm sat}/\mu \sim 7$ kV/cm (for electrons), 16 kV/cm (for holes)  -- conditions that are readily achieved for the terahertz pulses used in these experiments.

We used the split-step numerical method described above, together with the nonlinear Drude relations described in \eqref{eq:7}-\eqref{eq:9} to calculate the power-dependent transmission as a function of input power for the two waveguides under consideration.  For the p-doped silicon sample, we assumed a carrier concentration of $N = 8.5\times10^{13} $ cm$^{-3}$, a low-field hole mobility of $\mu = 470$ cm$^2/$(V$\cdot$s), a hole effective mass of $m^* = 0.36\, m_0$, and a saturation velocity of $0.75\times10^7$ cm/s.  For the high-resistivity float-zone silicon, we estimated a residual electron concentration of $N = 5\times10^{11} $ cm$^{-3}$ and a low-field electron mobility of $\mu = 1,416$ cm$^2/$(V$\cdot$s), effective mass of $m^* = 0.26\, m_0$, and an electron saturation velocity of $10^7$ cm/s.  In both cases the energy relaxation rate was taken to be $1/\Gamma_\varepsilon = 0.2$ ps.  We used accepted physical parameters from the literature, and the only adjustable parameter in the calculation was the carrier concentration $N$, which was chosen to both match the resistivity range of the wafers and to also agree with the observed absorption in the low-field limit.  The calculations were performed using 100 $\mu$m steps and a temporal window of 80 ps divided into 4000 steps.  For the numerical calculations, the input THz waveform was taken to be of the form $E(t) = E_0\cos(at)e^{-bt^2}$ where the constants $a$ and $b$ were chosen to best match the actual measured input waveform.  The dashed lines in Fig.~\ref{fig:3} show the calculated pulse energy transmission as a function of the input power and peak field, and agree well with the experimental measurements.

\begin{figure}[tbp]
\centering
\includegraphics{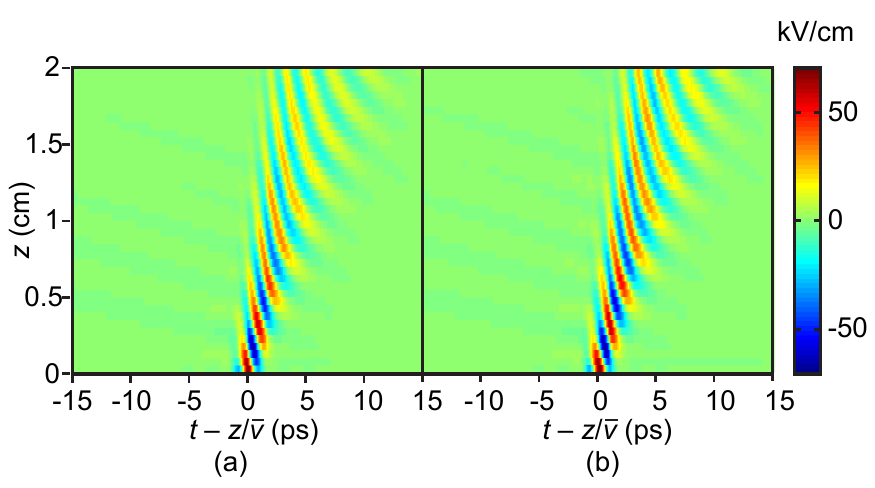}
\caption{Evolution of the temporal profile of terahertz pulse along the waveguide obtained from (a)conventional Drude model and (b)nonlinear split-step simulation}
\label{fig:4}
\end{figure}

Fig.~\ref{fig:4} shows a numerical simulation of how the terahertz pulse evolves in time as it traverses the 2 cm long waveguide.  The left portion was calculated using the conventional Drude model while the right portion includes the nonlinear split-step model discussed here, assuming a peak-peak input field of 100 kV/cm, clearly showing the enhanced field transmission.

\begin{figure}[tbp]
\centering
\includegraphics{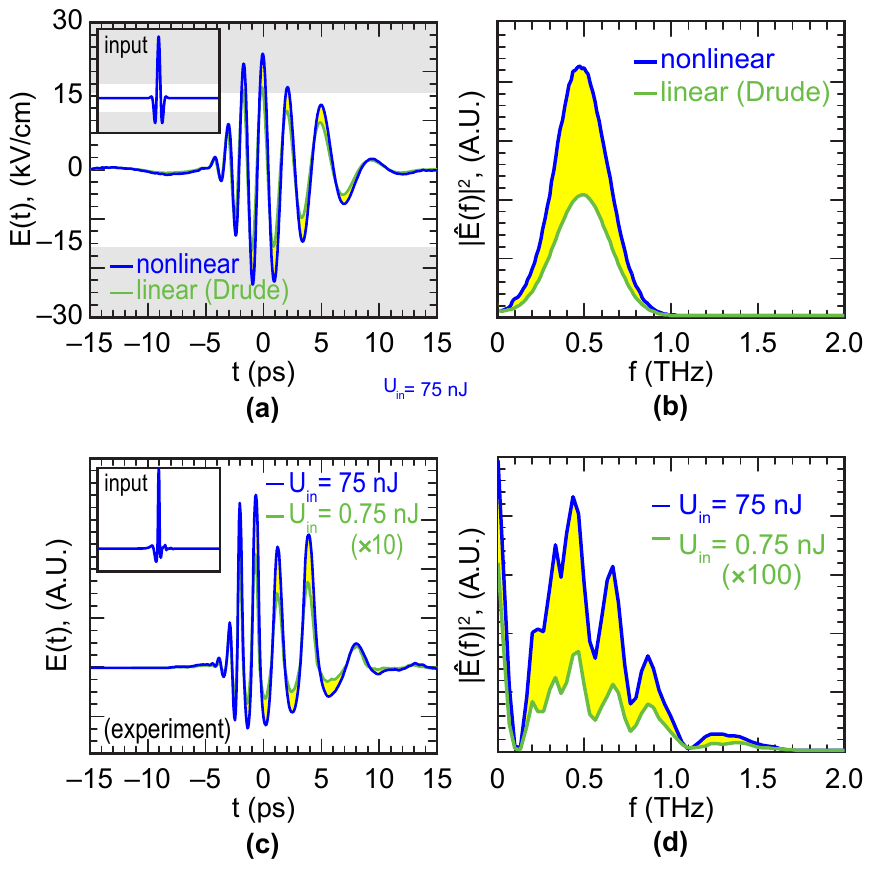}
\caption{(a) Transmitted terahertz waveform, calculated using Monte Carlo simulation of carrier dynamics together with the split-step Fourier method.  The linear (green) output curve was calculated using the conventional linear Drude model and waveguide dispersion, and shows lower transmission.  Inset: the simulated input pulse, with a peak-peak field of 100 kV/cm.  The Supplemental Material at [URL will be inserted by publisher] provides an animation showing how the field and nonlinearity evolve with distance. (b) Transmitted power spectrum, calculated with (blue) and without (green) nonlinearity.  (c) Experimentally measured transmitted terahertz waveform for 75 nJ (blue) and 0.75 nJ (green) incident pulse energy.  The green field was scaled by $10\times$ to account for the $100\times$ lower energy.  (d) Experimentally measured power spectra for 75 nJ (blue) and 0.75 nJ (green) incident pulse energy.  The green curve was scaled by $100\times$ to account for the lower energy.}
\label{fig:5}
\end{figure}

The balance equations \eqref{eq:7}-\eqref{eq:9} provide a simple and efficient model of the nonlinear transport in silicon, but there are alternative models used to explain the energy-dependent relaxation rates.  The most accurate and widely accepted approach is to use the Monte Carlo method to directly simulate the Boltzmann transport equations in the time domain\cite{lundstrom2002,*grasser2003,*willis2010}.  To better resolve the physical origins of the nonlinearity, we used the same split-step Fourier method to compute the nonlinear propagation, but instead of \eqref{eq:7}-\eqref{eq:9}, the current density at each step was estimated using time-dependent Monte Carlo simulations of an ensemble of 10,000 carriers.  This method is far more computationally intensive, and we therefore divided the waveguide into only 20 steps and simulated propagation for an input pulse with peak-peak field of 100 kV/cm. The same enhancement of transmission is observed (Fig.~\ref{fig:5}).

The Monte Carlo calculations incorporate several physical effects that contribute to the observed response, including band non-parabolicity, Coulomb scattering, intravalley acoustic phonon scattering, and equivalent  intervalley optical phonon scattering.  Of these, simulations revealed that intravalley and equivalent intervalley phonon scattering were found to be the dominant factors that contribute to the nonlinearity in the simulated response.  Notably, higher energy L-X intervalley scattering does not play a significant role, as had been previously suggested.

Fig.~\ref{fig:5}(a)-(b) show the calculated output waveform and spectra, obtained using a combination of the Monte Carlo simulation with split step Fourier method, for the highest input field (100 kV/cm) that was considered in the p-doped waveguide.  For comparison we also show the field obtained from the conventional (linear) Drude model, which would predict a higher carrier velocity and lower output field.  The Supplemental Material at [URL will be inserted by publisher] provides an animation showing how the nonlinearity and dispersion accumulate as the pulse traverses waveguide.

Fig.~\ref{fig:5}(c)-(d) show the corresponding experimental measurements of the output terahertz waveforms and spectra, which show a similar increase in relative transmission at high fields.  The experimental spectra show additional loss that is attributed to strong water absorption at 0.55, 0.75 and 1.1 THz that is absent from the simulations.  To assess the role of nonlinearity, we attenuated the input power by a factor of 100$\times$ and repeated the measurement of the output waveform.  The green linear curve shown in Fig.~\ref{fig:5}(c)-(d) was then scaled by a factor of 10 or 100 to provide a direct comparison with the field and power (respectively) measured at higher power.  These measurements clearly show that the relative transmission is significantly higher for strong terahertz pulses, and the degree of absorption saturation is comparable to that shown in Fig.~\ref{fig:5}(b).

In conclusion, we experimentally explore the phenomenon of absorption saturation in silicon dielectric waveguides at terahertz frequencies.  The field-induced transparency and associated carrier velocity saturation is shown to be dynamical effect that cannot be adequately explained by a modified effective mobility or Drude model.  We present a simple, nonlinear Drude model that explains the observations, and we confirm the model using rigorous Monte Carlo simulations.  Further, we introduce a numerical split-step method that models the interplay of nonlinearity and dispersion in the wave propagation.  These results could have important consequences in future high-power terahertz guided-wave nonlinear devices, such as terahertz frequency converters, parametric oscillators, mixers, and modulators.


\end{document}